\documentclass[%
 reprint,
 amsmath,amssymb,
 aps,
]{revtex4-2}

\usepackage{graphicx}% Include figure files
\usepackage{dcolumn}% Align table columns on decimal point
\usepackage{bm}% bold math
\usepackage{amsthm,amsmath,amssymb}
\usepackage{mathrsfs}
\begin{document}

\preprint{APS/123-QED}

\title{Measurement of Neutron Lifetime and Purcell Effect}% Force line breaks with \\
%\thanks{A footnote to the article title}%

\author{Fei He}
\author{Ka-Di Zhu}%
 \email{zhukadi@sjtu.edu.cn}
\affiliation{%
 Key Laboratory of Artificial Structures and Quantum Control (Ministry of Education), School of Physics and Astronomy, Shanghai Jiao Tong University,
800 Dong Chuan Road, Shanghai 200240, China
}%

%\collaboration{MUSO Collaboration}%\noaffiliation

\date{\today}% It is always \today, today,
             %  but any date may be explicitly specified

\begin{abstract}
Purcell effect predicts that spontaneous radiation is not an intrinsic property of matter, but is affected by the environment in which it is located, and is the result of the interaction of matter and field. Purcell effect can be inferred from Fermi's Gold rule through strict quantum electrodynamics (QED), and through it can achieve the enhancement or suppression of radiation. We suggest that, in principle, the Purcell effect can be detected at the percentage level of neutron decay in experiments with trapped ultra-cold neutrons. As a test of our claim, we propose a currently achievable experimental protocol that can detect whether Purcell effect has occurred in an trapped ultra-cold neutron lifetime measurement experiment. Finally, we discuss the discrepancy in current methods of measuring neutron lifetime, which may be caused by different experimental setups.

%\begin{description}
%\item[Usage]
%Secondary publications and information retrieval purposes.
%\item[Structure]
%You may use the \texttt{description} environment to structure your abstract;
%use the optional argument of the \verb+\item+ command to give the category of each item. 
%\end{description}
\end{abstract}

%\keywords{Suggested keywords}%Use showkeys class option if keyword
                              %display desired
\maketitle

%\tableofcontents

\section{Introduction}

The $\beta$ decay is a common form of radioactive decay and is an important process in the production of heavy elements in stars. So far it has be detected in almost all isotopes of the element, except for the most extreme heavy elements on the nuclide chart. In the framework of quantum mechanics, Fermi proposed a simple beta decay theory based on Pauli's neutrino hypothesis in 1934\cite{doi:10.1119/1.1974382}. Considering the time-dependent interaction between a radiating system and electromagnetic fields, Fermi made an analogy of $\beta$-decay with the emission of electromagnetic radiation by the nucleus. The essential feature of this decay can be derived from the basic expression for the transition probability of an interaction that is weaker than the interaction that forms quasi-stationary states. Fermi hypothesized that heavy particles (protons and neutrons) can be treated as two internal quantum states of a heavy particle, i.e., neutrons and protons form an isospin doublet state. From the perspective of perturbation theory, the transition probability from the initial state $\mathinner{|i \rangle}$ to the final state $\mathinner{|f \rangle}$ is given by the following equation, which is known as the Fermi's Gold rule: $\mathcal{W}=\frac{2\pi}{\hbar}| \mathcal{M}_{i\rightarrow f}|^2 \rho_f$, where $\rho_f$ is the final state density, and $\mathcal{M}$ is the transition matrix element between initial and final states. Although this theory is incomplete (for example, it does not permit parity violation), it can describe the electron energy spectrum of decay and give a qualitative understanding of the value of the decay half-life.

More generally, Fermi’s golden rule is a first-order approximation in the case of time-dependent perturbations, which is of great significance to the study of the transition of particles between different energy levels. In particular, the Fermi’s golden rule can be used to study the transition probability of spontaneous emission of two-level atoms in the excited state under the action of a vacuum field. In the framework of quantum electrodynamics, according to Fermi's golden rule, the spontaneous emission rate is not only related to the properties of the atoms, but also related to the electric field intensity corresponding to the position of atoms and the state density of photons. It is an effective method to strengthen or suppress the spontaneous emission of atoms by changing the density of the radiation field mode near the atomic transition frequency. Purcell proposed in 1946 that if an atom is in a microcavity, the rate of spontaneous emission will be changed\cite{PhysRev.69.37}. This idea is of great interest, and the phenomenon of changing the environment of the atom to regulate the spontaneous emission rate of the atom is called the Purcell effect. Purcell effect widely exists in metamaterials\cite{PhysRevB.87.035136,PhysRevB.92.195127}, acoustic systems\cite{PhysRevLett.120.114301} and various systems\cite{PhysRevLett.106.030502,PhysRevLett.123.153001,PhysRevB.96.235143,PhysRevB.94.134204,PhysRevLett.121.064301,PhysRevB.94.195314,doi:10.1021/acsnano.7b07402,https://doi.org/10.1002/adma.201904132,PhysRevApplied.11.051001} and is the beginning of cavity quantum electrodynamics (CQED), which will be described below.

As discussed above, the Purcell effect is widely present in various emitters, and we doubt whether this effect, which changes the decay rate of the emitter, will occur in the bottle-type neutron decay experiment, due to the existence of the trap bottle. Namely, the electromagnetic environment around the emitter (it refers to the neutron, in this case) is altered by the presence of the trap bottle, following Fermi's idea and thus changing the lifetime of the neutron. By following this idea, we will investigate in this paper whether it is possible to derive in detail the Purcell effect that occurs in the physical decay process of nuclei and particles, particularly bottle-type neutron decay experiments.

Indeed, neutrons play a very important role in the study of physics and the structure of matter, especially in nuclear physics and particle physics. Bound neutrons and protons together form the visible Universe. On the other hand, free neutrons have a lifetime of only about fifteen minutes, which decays through the $\beta^-$ decay $n\rightarrow p+e^-+\bar{\nu}_e$. $\beta$ decay occurs by exchanging $W$ bosons, so this is a weak interaction. The lifetime of neutrons can precisely reflect the coupling constant of weak interactions\cite{Abele}, even the Big-Bang primordial nucleosynthesis theory\cite{PhysRevD.71.021302}, which is an important physical quantity for testing the Standard Model and cosmology\cite{PhysRevD.88.073002}. 

In general, there are currently two different experimental techniques for measuring neutron lifetime. One method, also known as the beam method\cite{PhysRevLett.65.289,PhysRevLett.91.152302,PhysRevC.71.055502,PhysRevLett.111.222501}, extracts the lifetime of neutrons by capturing and counting protons produced by the decay of the neutron beam. In another method, ultracold neutrons (UCN) are trapped in a container, and after a period of time the number of remaining undecayed neutrons in the container is counted to fit the lifetime of the neutrons according to the exponential decay law\cite{PhysRevLett.63.593,PICHLMAIER2010221,PhysRevC.85.065503,KHARITONOV198998,PhysRevC.97.055503}. Especially the latter bottle type experiment is particularly interesting from the perspective of QED. As discussed in this paper, such experiments offer the possibility of testing whether Purcell effects affect the neutron lifetime of decay in a trap potential.

The main point of this paper is to suggest that the tiny Purcell effect may have an impact on the bottle-type neutron lifetime measurement experiment. Basically, the idea is that neutrons and protons are two different quantum states of the nucleus, and neutron decay is a process of transitioning to the proton state; On the other hand, the presence of a trap may modify the state density, causing the Purcell effect to be present in a bottle-type neutron decay experiment.

An interesting fact about neutron decay is worth mentioning, i.e., the results of the two types of decay experiments can be roughly divided into two classes. The lifetime measured by the beam method is longer, and two recent experiments give an average of $\tau_n^{beam}=888.0\pm 2.0$ s\cite{Byrne_1996,PhysRevLett.111.222501}. In addition, by employing the bottle method, the average lifetime measured is $\tau_n^{bottle}=879.5\pm 0.5$ s\cite{SEREBROV200572,PhysRevC.78.035505,Ezhov,ARZUMANOV201579,PhysRevC.97.055503,Pattie627}. These results show that the experimental measurement values $\tau_n$ for the bottle-type are about $\sim$9 seconds less than those for the beam-type experiments, in other words, the neutrons in the trap will decay faster.

Up until recently, this non-negligible disagreement of about 9 seconds between the above two experimental scenarios is known as a "neutron decay anomaly" and still confuses people\cite{RevModPhys.83.1173}. On the one hand, this confusion has inspired experimental physicists to refine their experimental schemes to eliminate possible unknown errors\cite{Serebrov,10.1093/ptep/ptz153,wei2020new,hirota2020neutron}. One of the interesting schemes is that Wilson \emph{et al.} used data from NASA’s MESSENGER spacecraft during the flyby of Venus and Mercury to estimate the neutron lifetime to be approximately $780\pm 60(stat)\pm 70(sys)$ seconds\cite{PhysRevResearch.2.023316}. In addition, theorists have developed many theoretical investigations to explain this discrepancy. One of the efforts is that in the beam current experiment, when protons are stored in the magnetic trap, the protons are lost due to the charge exchange collision between the protons and the remaining gas\cite{serebrov2020possible,wietfeldt2020comment}. Another interesting idea is to take into account the inverse quantum Zeno effect (IZE), i.e., the observation of a quantum system, under certain conditions, will speed up its decay. The authors suggest that the inverse quantum Zeno effect is, in principle, detectable in neutron decay experiments\cite{PhysRevD.101.056003}. 

In addition, more interesting, Fornal and Grinstein proposed that one possible reason for the problem of neutron lifetime is that neutrons may decay into invisible dark matter particles\cite{PhysRevLett.120.191801}. Following this idea, Cline and Cornell explored the decay of neutrons into dark Dirac fermions $\chi$ and dark photons $A^{\prime}$\cite{Cline}. If the dark decay of neutrons is possible, then it should also occur in nuclei with lower binding energy for neutrons. Pf{\" u}tzner and Riisager argue that since the channel $^{11}Be(\beta p)$ is a quasi-free neutron decay mode\cite{RIISAGER2014305}, the neutron-rich nucleus $^{11}$Be is the most promising one to observe the dark decay of neutrons\cite{PhysRevC.97.042501}. However, both two experimental groups exclude the possibility of $n\rightarrow \chi +\gamma$ decay mode with 97\% confidence and a one percent contribution for 95\% of the $n\rightarrow \chi +e^+e^-$ decay mode\cite{PhysRevLett.121.022505,PhysRevLett.122.222503}. Meanwhile, Dubbers \emph{et al.} argue that the exotic decay mode is not the cause of the neutron lifetime anomaly with a high level of confidence\cite{DUBBERS20196}.

Another intriguing theory, the mirror matter theory, suggests that all ordinary particles have corresponding mirror partners, which could help explain the dark matter puzzle. This idea has recently been used to explain the neutron lifetime anomaly\cite{Berezhiani,TAN2019134921,Zurab}. The theory indicates that about 1\% fraction of ordinary neutrons are converted into mirror neutrons, which also undergo $n^{\prime}\rightarrow p^{\prime}e^{\prime}\bar{\nu}^{\prime}_{e}$ decay. What is particularly interesting is that under the new theory of Ref.\cite{TAN2019134921}, neutron oscillation becomes the messenger of the two worlds, allowing ordinary and mirror worlds, which otherwise have no other interactions, to exchange matter.

In summary, the neutron decay anomaly is more likely to be attributed to the systematic effects of the two experiments, as discussed in Ref.\cite{PhysRevLett.120.191801,PhysRevLett.120.202002,DUBBERS20196,PhysRevLett.122.242501}. In any case, exploring new experimental methods for measuring neutron lifetime will help to further understand the discrepancy in current experiments.

In the present case, one can explore a speculative and interesting idea: it is possible to assume that the tiny Purcell effect might have already occurred in the ongoing bottle-type experiments. In this way, the Purcell effect may provide a possible explanation for neutron decay anomaly because, for this particular experimental setup, shorter lifetimes are the result of increased decay rates.

\section{Brief Review of the CQED and Purcell effect}

In this section, we present a theoretical description of Purcell effect. For this purpose, we use the results given in Ref.\cite{Walther_2006}, which describe the theoretical model we will use in our approach.

In short, the Purcell effect is to increase the probability transition of a particle by changing the density of the state, thus increasing the decay rate of the particle. A more detailed explanation of the Purcell effect will be discussed in detail later. At present, in order to understand the underlying physical mechanism in detail, we might as well start with the spontaneous emission of atoms. The so-called spontaneous emission refers to the process in which an electron can only stay on the excited energy level of an atom for a short period of time, then spontaneously jumps to a lower energy level and radiates a photon. 

As we mentioned in the introduction, the spontaneous emission problem can be simply derived from the Fermi's golden rule, i.e., the transition probability of atomic spontaneous emission
\begin{equation}\label{eq1}
    W=\frac{2\pi}{\hbar^2}|M_{12}|^2g(\omega),
\end{equation}
where, $M_{12}=\mathinner{\langle}p\cdot\epsilon\mathinner{\rangle}$ is the transition matrix element of a two-level atom, and $g(\omega)=\omega^2V_0/\pi^2c^3$ is the photon density of states in free space. The mean lifetime $\tau$ of the system is related to $W$ by $\tau=1/W$. Considering that the polarization directions of all atoms are arbitrary relative to the direction of the field, it is necessary to average the transition matrix elements. In terms of vacuum$\epsilon_{vac}=(\hbar\omega/2\epsilon V)^{1/2}$, we can go to
\begin{equation}
    M_{12}^2=\frac{1}{3}\mu_{12}^2\epsilon_{vac}^2=\frac{\mu_{12}^2\hbar\omega}{6\epsilon_0V_0}.
\end{equation}
And considering Eq.(\ref{eq1}), we can finally deduce the transition probability as $W_{free}=1/\tau_R=\mu_{12}^2\omega^3/3\pi\epsilon_0\hbar c^3$. This coefficient is actually known as the Einstein $A_{21}$ coefficient.

Historically, Einstein proposed the $A$ and $B$ coefficients, which seemed to have successfully explained the problem of spontaneous emission rate of atoms, but in fact, this problem has not been completely solved. A fundamental and significant question is why atoms in an excited state "spontaneously" fall to the ground state. Actually when we calculated the transition matrix elements above, we have already considered the interaction between the two-level atom and the vacuum. In other words, it is precisely the coupling of free space with the emitter that causes spontaneous emission, and this problem has been solved within the framework of QED. Later, the emitter will be defined as a neutron, but here it refers to a two-level atom.

In addition, in the calculations above, we consider the interaction between atoms and vacuum. A natural idea is whether to change the properties of space and to modify the lifetime of atoms at a certain energy level. Purcell realized that according to the golden rule, the spontaneous emission rate of an atom is related to its environment. For example, in a resonator cavity, because of the boundary conditions provided by the cavity wall, the spectrum of the cavity mode changes compared to the case in free space, i.e., the density of the state changes. Thus, the spontaneous emission of atoms can be enhanced or suppressed by controlling their environment. 

For simplicity, we consider that only one resonance mode $\omega_c$ of the cavity interacts with the atom, and the other modes of the cavity are far away from the resonance frequency of the atom. The cavity mode density satisfies the normalized condition $\int_0^{\infty}g(\omega)d\omega=1$, so one can arrive the mode density of the normalized Lorentz spectrum in the cavity as
\begin{equation}
    g(\omega_A)=\frac{2}{\pi\Delta\omega_c}\frac{\Delta\omega_c^2}{4(\omega_A-\omega_c)^2+\Delta\omega_c^2},
\end{equation}
where $\Delta\omega_c$ and $\omega_A$ are the half-width of the spectral line and the transition frequency of the atom, respectively. By introducing the normalized dipole orientation factor $\zeta=|p\cdot\epsilon|/|p|\cdot|\epsilon|$, the transition matrix element can be written as $M_{12}^2=\zeta^2\mu_{12}^2\epsilon_{vac}^2$, and therefore the transition probability of the atoms in the cavity can be deduced as
\begin{equation}
    W_{cav}=\frac{2Q\zeta^2\mu_{12}^2}{\hbar\epsilon_0 V_0}\frac{\Delta\omega_c^2}{4(\omega_A-\omega_c)^2+\Delta\omega_c^2},
\end{equation}
where $Q=\omega_c/\Delta\omega_c$ is the quality factor. This result is different from the previously discussed transition probability of atoms in free space, and it shows that the existence of the cavity will change the state density, thereby modifying the transition probability of the emitter. In order to characterize the effect of the presence of cavity on the transition probability of the emitter, Purcell factor $F_p
\equiv W_{cav}/W_{free}$ is defined: if $F_p$ is greater than 1, the spontaneous emission of the emitter is enhanced; on the contrary, if $F_p$ is less than 1, it is suppressed. 

The proposal of Purcell effect became the beginning of cavity quantum electrodynamics. The existence of a cavity will change the vacuum, and cavity quantum electrodynamics mainly studies the matter and field in a specific confined space, including the space itself and the quantum phenomena of interaction. In this paper, our main idea is that in future bottle-type experiments, analogous to the Purcell effect in CQED, the existence of a trap may change the nature of the vacuum, which may play an important role in experiments designed to measure the neutron lifetime. In the next section, we will inherit the view that "the decay of a neutron to form a proton is equivalent to a transition between two quantum states", and consider the existence of a bottle to change the vacuum, showing how the bottle-type experiment modifies the neutron decay.

\section{The Purcell Effect and the neutron: Qualitative features}

Let us discuss the Purcell effect of neutron decay. First, the Hamiltonian describing the $\beta$ decay of the nucleus is written as:
\begin{equation}
    \mathcal{H}=H_{nuc}+H_{e}+H_{\nu}+H_{W}.
\end{equation}
The four terms that constitute the total Hamiltonian represent the initial and final nuclei, electron, neutrino, and weak interaction. Considering the $V–A$ standard electroweak model, the matrix element of the Hamiltonian for allowed $\beta$ decay can be written as 
\begin{equation}
    H_{W}^{V-A}=2g[\bar{p}(C_V-C_A\gamma_5)\gamma_{\mu}n](\bar{e} \gamma_{\mu}\nu)+h. c. ,
\end{equation}
where $g$ is the universal coupling constant, $p$, $n$, $e$ and $\nu$ are the wave functions of proton, neutron, electron and neutrino respectively, and $C_A$ and $C_V$ are the coupling constants of vector and pseudovector respectively. $\gamma_5$ and $\gamma_{\mu}$ are both Dirac matrices. Physically, although the theory is non-renormalizable, this quantity determines the width $\Gamma_{\beta}=\Gamma_n$ of $\beta$ decay $n\rightarrow pe\bar{\nu}_e$. In this context, we take the view that the tiny Purcell effect can be seen as a perturbation in future bottle-type neutron lifetime measurement experiments. In this case, the beam method should measure the true decay width $\Gamma_{beam}=\Gamma_n=\tau_{n}^{-1}$ of the neutron, whereas in the bottle method one actually measures $\Gamma_{bottle}=\Gamma_{n}+\Gamma_{P}$. Where, $\Gamma_{P}$ is the extra decay width caused by the Purcell effect, which accelerates the decay rate of the neutron. This hypothesis implies that the neutron lifetime given by the bottle method is shorter than that measured by the beam method, possibly explaining 1\% of the experimental inconsistencies. It should be noted that this paper is only qualitatively suggesting that in ongoing bottle-type neutron lifetime experiments, one might assume that the Purcell effect has occurred. In this case, a possible explanation for the neutron lifetime puzzle may be provided by the special setup of the experimental apparatus.

\begin{figure}[b]
\includegraphics[width=0.45\textwidth]{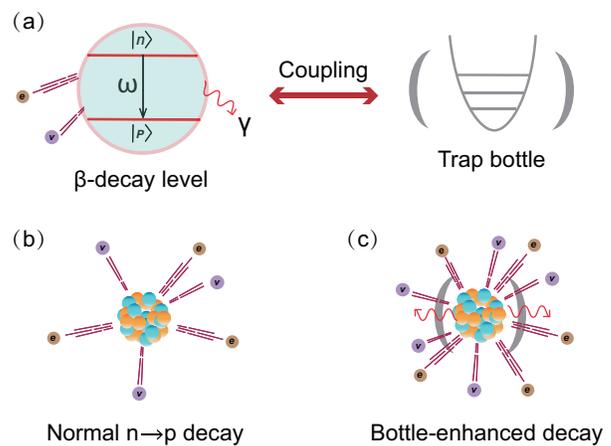}% Here is how to import EPS art
\caption{\label{fig:epsart} Schematic diagram of the interaction our proposed between the neutron and the bottle that has not taken into account before in the bottle-type neutron decay experiment. (a) We herein regard the decay $n\rightarrow pe\bar{\nu}_e$ as a transition between the $\mathinner{|n \rangle}$ and $\mathinner{|p \rangle}$ states of a pseudo-two-level system, and the trap bottle is equivalently a cavity. Since there may be energy exchange between the neutron and cavity, we can explore the influence of coupling on neutron lifetime measurement in the framework of CQED. (b) Schematic diagram of free neutron decay. (c) The neutron decay is enhanced due to the interaction between the neutron and the cavity, which in turn leads to a smaller measured value in the bottle-type experiment.}
\end{figure}

Since there is a mass difference of $M_n-M_p\simeq+1.29MeV/c^2$ between proton and neutron\cite{Olive_2016}, in the case of our hypothesis, we equivalently regard neutrons and protons as an effective pseudo-two-level system, and the Hamiltonian describing the system can be written as $\hbar\omega\sigma_z$, where $\omega$ is the $neutron\rightarrow proton$ resonant frequency and $\sigma_z=\mathinner{|n \rangle}\mathinner{\langle n|}-\mathinner{|p \rangle}\mathinner{\langle p|}$ is the pseudospin operator. Physically, $\mathinner{|n \rangle}$ and $\mathinner{|p \rangle}$ can be regarded as the two eigenstates of the Pauli operator $\sigma_z$, whose eigenvalues correspond to $+1$ and $-1$, respectively. In fact, the form of the Hamiltonian to be diagonalized is the same as the form of the Hamiltonian for the spin $1/2$ particle in the static magnetic field $\boldsymbol{B}$. In other words, one can regard any two-level system as a particle with a spin (pseudo-spin) of 1/2 in a static magnetic field $\boldsymbol{B}$. This two-level system can be represented by the leftmost ``energy level diagram" in Fig.~\ref{fig:epsart}(a). The symbols $\mathinner{|p \rangle}$ and $\mathinner{|n \rangle}$ represent the proton state and the neutron state, respectively. In fact, the simplest complex matrix group that describes the transformation between $\mathinner{|p \rangle}$ and $\mathinner{|n \rangle}$ states and satisfies the conservation of probability is the SU(2) group. In the following, we immediately realize that the neutron-proton transition depicted by the SU(2) group can be matched to a classical model of cavity quantum electrodynamics, so from a symmetry point of view we make an analogy to explore the Purcell effect that may have occurred in ongoing neutron decay experiments.

On the other hand, let us consider the influence of the experimental setup on the vacuum state of the environment in which the neutron is located. The $k$-th mode of this bottle field has a resonant frequency of $\omega_k$, and the quantized Hamiltonian can be written as $H_{b}=\sum_k \hbar\omega_k a_k^{\dagger}a_k$, which is commonly used in quantum optics. Operators $a_k^{\dagger}$ and $a_k$ are the creation and annihilation operators of bottle field respectively, and they satisfy the commutation relation $[a_k, a_k^{\dagger}]=1$.

For mathematical simplicity, we consider that under the dipole approximation, the interaction between the bottle and its contents is $H_{de-tb}=-e\vec{r}\cdot \vec{E}=g(a^{\dagger}+a)(\sigma_++\sigma_-)$, where $g$ characterizes the coupling strength of this interaction and in the two-level case, $\sigma_+$ as well as $\sigma_-$ are Pauli operators respectively:  
\begin{equation}
    \begin{aligned}
        \sigma_+=\begin{pmatrix} 0 & 1 \\ 0 & 0 \end{pmatrix},
        \sigma_-=\begin{pmatrix} 0 & 0 \\ 1 & 0 \end{pmatrix}.
    \end{aligned}
\end{equation}
The decay system should satisfy the conservation of energy. Therefore, under the rotating wave approximation, one can rewrite $H_{de-tb}$ as $H_{int}=\hbar \sum_k g_k(\sigma_-a_k^{\dagger}+\sigma_+a_k)$. Physically, the first and second terms represent probably $\mathinner{|n \rangle}\rightarrow \mathinner{|p \rangle}$ and $\mathinner{|p \rangle}\rightarrow \mathinner{|n \rangle}$ transitions, respectively. Physically, we realize that this coupling is probably caused by the energy exchange between the neutron and the trap bottle.

Now then, we can infer that if nucleons interact with the experimental setup, as shown in the above analysis, the bottle-type experiment would accelerate the decay rate of neutrons, possibly explaining the current lifetime puzzle. In the following, we follow the approach in Ref. \cite{PhysRev.69.37}. Within this method, the perturbation Hamiltonian describing the bottle-type experiment system is given by
\begin{equation}\label{1}
    \mathscr{H}=\sum_{k}\hbar \omega_k a_k^{\dagger}a_k+\hbar\omega\sigma_z+\sum_{k}g_k(\sigma_-a_k^{\dagger}+\sigma_+a_k).
\end{equation}
In order to study the decay dynamics of a pseudo-two-level system composed of proton and neutron, we put the operator $\sigma_z$ into the Heisenberg equation $i\hbar \frac{d\hat{O}}{dt}=[\hat{O},\mathscr{H}]$, and taking into account the commutative relation $[\sigma_z,\sigma_{\pm}]=\pm\sigma_{\pm},[\sigma_+,\sigma_-]=2\sigma_z$ as well as $\{\sigma_+,\sigma_-\}=\mathcal{I}$, one can obtain:
\begin{equation}\label{2}
    \dot{\sigma}_z=-i\sum_k g_k(a_k\sigma_+-a_k^{\dagger}\sigma_-).
\end{equation}
The same mathematical technique applies to the $a_k\sigma_+$ and $a_k^{\dagger}\sigma_-$ operators, we have
\begin{equation}
 \begin{aligned}
     \frac{d}{dt}(a_k\sigma_+)=&i(\omega-\omega_k)a_k\sigma_+-ig_k\left(\sigma_z+\frac{1}{2}\right) \\ &-2i\sum_{k^{\prime}}g_{k^{\prime}}\sigma_z a_{k^{\prime}}^{\dagger} a_k,  \\
    \frac{d}{dt}(a_k^{\dagger}\sigma_-)=&-i(\omega-\omega_k)a_k^{\dagger}\sigma_-+ig_k\left(\sigma_z+\frac{1}{2}\right) \\
    &+2i\sum_{k^{\prime}}g_{k^{\prime}}\sigma_z a_{k^{\prime}}a_k^{\dagger}.
 \end{aligned}
\end{equation}
Integrate the above two equations respectively, and we assume that at time $t=0$, the radiation field is in a vacuum state, one can write
\begin{equation}
    \begin{aligned}
        a_k\sigma_+(t)=&-ig_ke^{i(\omega-\omega_k)t}\int_0^te^{-i(\omega-\omega_k)\tau}d\tau   \\
        &\times\left[\sigma_z(\tau)+\frac{1}{2}\right]+a_k\sigma_+(0)e^{i(\omega-\omega_k)t}, \\
        a_k^{\dagger}\sigma_-(t)=&ig_ke^{-i(\omega-\omega_k)t}\int_0^te^{i(\omega-\omega_k)\tau}d\tau   \\
        &\times\left[\sigma_z(\tau)+\frac{1}{2}\right]+a_k^{\dagger}\sigma_-(0)e^{-i(\omega-\omega_k)t}.
    \end{aligned}
\end{equation}
Substituting the above results into Eq.(\ref{2}), and considering that the neutron decay time is about fifteen minutes, the application of Markov approximation leads to
\begin{equation}
    \frac{d}{dt}\langle\sigma_z(t) \rangle=-\left[\langle\sigma_z(t) \rangle+\frac{1}{2}\right]\cdot 2\pi\sum_kg_k^2\delta(\omega_k-\omega).
\end{equation}
If we set the parameter $\Gamma=2\pi\sum_kg_k^2\delta(\omega_k-\omega)$, then we can finally arrive the evolution law of the nuclear system
\begin{equation}
    \langle\sigma_z(t) \rangle=\left[\langle\sigma_z(0) \rangle+\frac{1}{2}\right]\rm{exp}(-\Gamma t)-\frac{1}{2},
\end{equation}
from which we can infer that the system decays exponentially, and that $\Gamma^{-1}$ is the decay lifetime. We recall $\Gamma$ and apply the transformation $\sum_k\rightarrow \frac{2V}{(2\pi)^3}\int dk^3$, and eventually we get to $\Gamma=2\pi g^2(\omega) \rho(\omega)$, where $g(\omega)=|d_{ij}|\xi(\omega)/\hbar$ and density of states $\rho(\omega)=V\omega^2/(\pi^2c^3)$. In the above equation, $\xi(\omega)=\sqrt{\hbar\omega/2\epsilon_0 V}$, $d_{ij}=e\mathinner{\langle i|} \textbf{r} \mathinner{|j \rangle}$ is the matrix element for the dipole transition, $V$ is the mode volume of the bottle radiation field and $\epsilon_0$ is the permittivity of vacuum. From this way we can infer that the decay rate of the bottle-type experiment can be adjusted, i.e., it can be achieved by adjusting the state density around the transition frequency. In other words, one may be able to control the lifetime of nuclear decay by changing the environment in which the neutron is located in the trap bottle.

\begin{figure}[b]
\includegraphics[width=0.45\textwidth]{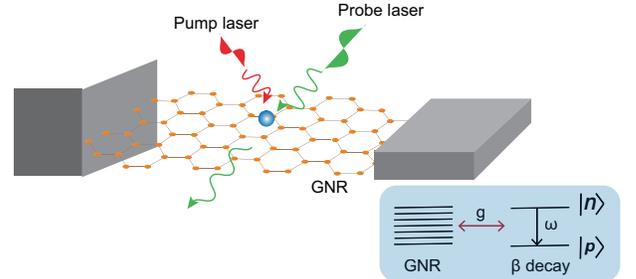}% Here is how to import EPS art
\caption{\label{Fig_2} Schematic of proposed neutron and GNR coupling system. The two ends of the nanoribbon are fixed, and neutrons are attached to the GNR. The whole system is driven by a strong pump pulse and a weak probe pulse. The inset is an energy levels diagram of the proposed pseudo-two-level system when dressing the eigenmode of GNR.}
\end{figure}

\begin{figure*}
\includegraphics[width=1\textwidth]{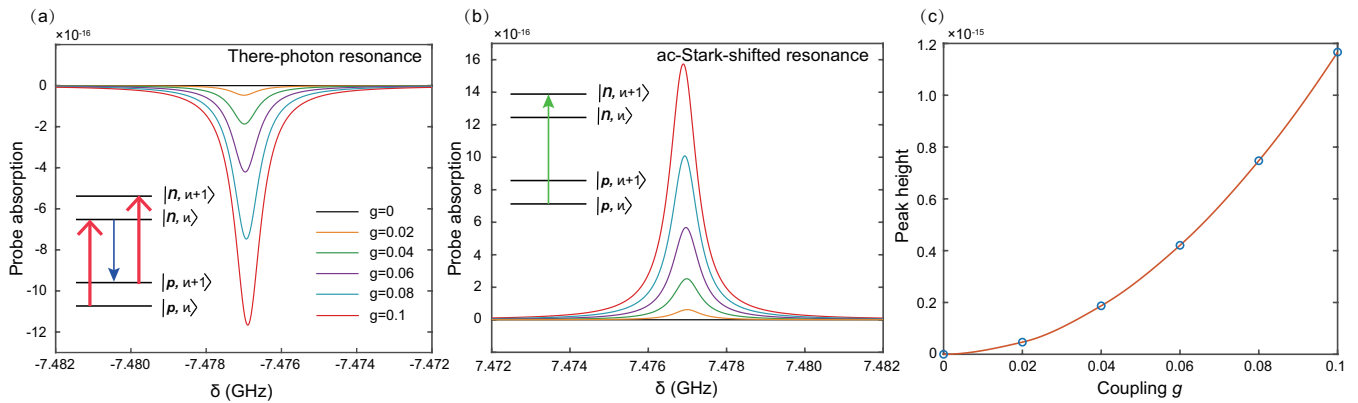}% Here is how to import EPS art
\caption{\label{fig:wide} Probe absorption spectrum as a function of detuning $\delta$ with different neutron-GNR coupling strength. (a) The peak is located at $\delta=-\omega_k$. We suggest that this is a there-photon resonance, and the inset shows the energy level diagram of this process. The parameter we choose here is $\Omega=10$ GHz, $\Delta_{pu}=100$ GHz, and $\omega_k=7.477$ GHz\cite{Sadeghi_2010}. (b) is similar to (a) except that the peak is located at $\delta=\omega_k$. We suggest that this is ac-Stark-shifted resonance. Both figures a and b show that as the neutron GNR coupling increases, the peak becomes higher, which is illustrated in detail in (c). The variation of peak height caused by coupling between neutron and GNR can be well recognized in the spectrum.}
\end{figure*}

In fact, experiments have shown that the decay of the nucleus may be affected by the environment. One interesting issue is that this decay rate enhancement effect on the nucleus may have been demonstrated in earlier experiments. Ohtsuki \emph{et al.} reported in 2004 that they experimentally measured the decay rates of $^7$Be electron capture (EC) $\beta$ decay in C$_{60}$ and Be metal, respectively\cite{PhysRevLett.93.112501}. They pointed out that there is a 0.83\% difference in the decay rate of $^7$Be in two different environments, and the rate enhances in the case of $^{7}$Be@C$_{60}$. Further experiments showed that they measured decay rate of $^7$Be in C$_{60}$ at the temperature $T=5K$ and found that low temperature can speed up decay\cite{PhysRevLett.98.252501}. The authors interpreted the observations by calculating the electron density at the position of $^7$Be inside C$_{60}$. Based on these experiments, Morisato \emph{et al.} used density functional theory to calculate and explain why $^7$Be@C$_{60}$ shows higher EC decay rate than $^7$Be metal found in the above experiments in detail, and found that there are inequivalent four stable Be sites inside C$_{60}$ and that center of C$_{60}$ is the most favorable site\cite{PhysRevB.78.125416}. Recently, a similar analysis was published in 2012\cite{PhysRevC.86.014608}. Tkalya \emph{et al.} calculated EC $\beta$ decay of $^7$Be when $^7$Be is encapsulated in C$_{36}$ cage in terms of \emph{ab initio} post-Hartree-Fock method and demonstrated that the chemical environment did affect nucleus $\beta$ decay properties. The above examples all show that the properties of spontaneous decay of atomic nuclei can be changed by the external environment. However, the mechanism of the above-mentioned experiments is completely different from the Purcell effect that may occur in the neutron $\beta^-$ decay experiment proposed in this paper.

As mentioned above, in our scheme, the effect of the decay rate enhancement can be measured in terms of the Purcell factor $Fp$, which is defined as the ratio of the enhanced decay rate to the spontaneous decay rate in vacuum. We propose that the neutron lifetime measured by the beam method is true lifetime, i.e., the spontaneous neutron decay lifetime in a vacuum and the measured lifetime in the bottle-type experiment is enhanced through the CQED effect. In this case, one obtain $\Gamma^{beam}=1.126\times 10^{-3}s^{-1}$ and therefore $Fp\approx 1.0097$. It is worth noting that in the experiment, CQED has been used to projective measurement of a single nuclear spin\cite{PhysRevA.81.062308,PhysRevLett.106.160501}.

In fact, a typical model describing the interaction between a two-level system with a quantum cavity field is the well-known Jaynes-Cummings model. The reason why the Jaynes-Cummings model is analytic is that it has an intrinsic SU(2) symmetric structure, so the model can always be converted into a 1/2-spin particle system under an effective static magnetic field. On the other hand, in Fermi's earlier model, i.e., the model adopted in this paper, the simplest complex matrix group for the transformation between a neutron state and a proton state that satisfies the conservation of probability is the SU(2) group. In this case, the decay of neutrons can be seen as following the same mechanism as the Jaynes-Cummings model.

\section{The Purcell Effect in future and ongoing neutron experiments: General considerations}

The existing theoretical and experimental studies have found that the decay properties of nucleus can be changed by the external environment\cite{PhysRevLett.93.112501,PhysRevLett.98.252501,PhysRevB.78.125416,PhysRevC.86.014608}, which potentially imply that the cavity quantum electrodynamics effect will affect the $n\rightarrow p+e^-+\bar{\nu}_e$ decay process in the bottle-type experiment. The main question in the Purcell effect of neutron decay is how to observe the modification of the neutron lifetime measurement by the experimental apparatus. This problem can be explored by the interaction between neutrons and experimental setup-an equivalent cavity.

However, to the best of our knowledge, there is no direct experimental evidence for Purcell enhancement in neutron $\beta^-$ decay. In this context, we also theoretically propose an experiment that can realize and detect the coupling of a neutron to an equivalent cavity. The schematic diagram of our proposed experiment is shown in Fig.~\ref{Fig_2}. One can couple neutrons with graphene nanoribbons (GNR), and the eigenmode of GNR is regarded as an equivalent cavity with a Hamiltonian of $\hbar\omega_ka^{\dagger}a$, where $\omega_k$ is the eigenmode frequency of the GNR. It must be emphasized that GNR used here, from a phenomenological point of view, is physically equivalent to a trap bottle or cavity. On the other hand, as mentioned above, a series of experiments and theoretical calculations show that the chemical environment of the nucleus, i.e., its position, will affect its decay properties\cite{PhysRevLett.93.112501,PhysRevLett.98.252501,PhysRevB.78.125416,PhysRevC.86.014608}. Therefore, one can regard the $\mathinner{|n \rangle}\rightarrow \mathinner{|p \rangle}$ transition frequency as being modulated by the GNR and rewrite it as
\begin{equation}
    \omega(x)\approx\omega+x\frac{\partial\omega}{\partial x}+\cdots. 
\end{equation}
Recalling the Hamiltonian describing neutron decay, we expand the transition frequency to the leading order of the displacement, and obtain
\begin{equation}
    \begin{aligned}
        \hbar \omega (x)\sigma_z &\approx\hbar (\omega-x\frac{\partial\omega}{\partial x})\sigma_z    \\
        &=\hbar\omega\sigma_z-\hbar x_{{\rm ZPF}}\frac{\partial\omega}{\partial x}(a^{\dagger}+a)\sigma_z.
    \end{aligned}
\end{equation}
Where $x=x_{{\rm ZPF}}(a^{\dagger}+a)$ and $x_{{\rm ZPF}}=\sqrt{\hbar/2m_{{\rm eff}}\omega_k}$ is the zero-point fluctuation of the GNR, $m_{{\rm eff}}$ is the effective mass. If we define the coupling strength as $g=x_{{\rm ZPF}}\frac{\partial\omega}{\partial x}$, the Hamiltonian of the interaction can be read as
\begin{equation}
    \mathcal{H}_{int}=-\hbar g(a^{\dagger}+a)\sigma_z.
\end{equation}

In order to detect the interaction between neutron and GNR, we introduce one pump beam to drive the whole system and another probe beam to detect, which can be described by
\begin{equation}
    \begin{aligned}
        H_{op}=&-\mu(\sigma_+E_{pu}e^{-i\omega_{pu}t}+\sigma_-E_{pu}^*e^{i\omega_{pu} t})   \\
        &-\mu(\sigma_+E_{pr}e^{-i\omega_{pr}t}+\sigma_-E_{pr}^*e^{i\omega_{pr} t}).
    \end{aligned}
\end{equation}
We recall the Eq.(\ref{1}) of the bottle-type experiment and obtain the following Hamiltonian in the frame of rotating wave transform that describes our proposed experiment\cite{PhysRevLett.92.075507}
\begin{equation}
    \begin{aligned}
        \mathfrak{H}=&\hbar\Delta_{pu}\sigma_z+\hbar\omega_k a^{\dagger}a -\hbar g\sigma_z(a^{\dagger}+a)  \\
        &-\mu(\sigma_+E_{pr}e^{-i\delta t}+\sigma_-E_{pr}^*e^{i\delta t})  \\
        &-\hbar(\Omega\sigma_++\Omega^*\sigma_-),
    \end{aligned}
\end{equation}
where $\Delta_{pu}=\omega-\omega_{pu}$ is the detuning of pump beam and $neutron\rightarrow proton$ resonant frequency, $\Omega=\mu E_{pu}/\hbar$ is the Rabi frequency of pump field, and $\delta=\omega_{pr}-\omega_{pu}$ is the detuning of probe beam and pump beam. We define $N=a^{\dagger}+a$ and substitute it with $\sigma_z$ and $\sigma_-$ into the Heisenberg equation, which follows the standard quantum optics procedure to arrive at the Following Langevin equation:
\begin{equation}
    \begin{aligned}
        \dot{\sigma}_z=&i\Omega(\sigma_+-\sigma_-)+\frac{i\mu}{\hbar}(E_{pr}e^{-i\delta t}\sigma_+-E_{pr}^*e^{i\delta t}\sigma_-) \\
        &-\Gamma_1\left(\sigma_z+\frac{1}{2}\right),
    \end{aligned}
\end{equation}
\begin{equation}
    \begin{aligned}
        \dot{\sigma}_-=&-(i\Delta_{pu}+i\omega_kgN+\Gamma_2)\sigma_--2i\Omega\sigma_z   \\
        &-\frac{2i\mu}{\hbar}E_{pr}e^{-i\delta t}\sigma_z,
    \end{aligned}
\end{equation}
\begin{equation}
    \begin{aligned}
        \frac{d^2N}{dt^2}+\gamma_k\frac{dN}{dt}+\omega_k^2N+2\omega_k^2g\sigma_z=0.
    \end{aligned}
\end{equation}
Here, $\Gamma_1$ and $\Gamma_2$ are phenomenologically introduced as the relaxation rate and dephasing rate of this pseudo-two-level system composed of neutrons and protons, respectively. In addition, $\gamma_k$ should be introduced to be the decay rate of the cavity due to the interaction between the cavity and the background modes\cite{PhysRevLett.92.075507}. In order to solve the above equations, we focus on the ansatz $\hat{O}=\hat{O}^0+\hat{O}^+e^{-i\delta t}+\hat{O}^-e^{i\delta t}~({\rm here}~\hat{O}=\sigma_z,\sigma_- ~{\rm and} ~ N)$, and then the steady-state solution of the system can be obtained. Furthermore, the dimensionless first-order optical susceptibility related to the probe field reads
\begin{equation}
    \chi(\omega_{pr})=\sigma_+^-\frac{\Gamma_2\hbar}{\mu E_{pr}}=\frac{2\varepsilon \eta_0 (\Omega_R^2+\zeta)-\varphi\eta_0}{\varpi\varphi-2\varepsilon(\Omega_R^2+\zeta)(\varepsilon-\delta_0)},
\end{equation}
the auxiliary functions are
\begin{equation}
    \begin{aligned}
       & \varpi=\Delta_{pu0}-\omega_{k0}g^2\eta_0-i-\delta_0,\\
        & \varepsilon=\Delta_{pu0}-\omega_{k0}g^2\eta_0+i+\delta_0, \\
        &  \varphi=(2\Omega_R^2+2\iota-2i\varepsilon-\varepsilon\delta_0)(\varepsilon-\delta_0), \\
        &  \iota=\frac{\Omega_R^2\omega_{k0}g^2\varrho\eta_0}{\Delta_{pu0}-\omega_{k0}g^2\eta_0+i}, \\
         & \zeta=\frac{\Omega_R^2\omega_{k0}g^2\varrho\eta_0}{\Delta_{pu0}-\omega_{k0}g^2\eta_0-i}, \\
         & \varrho=\frac{\omega_{k0}^2}{\omega_{k0}^2-\delta_0^2-i\delta_0\gamma_{k0}},
    \end{aligned}
\end{equation}
where $\delta_0=\delta/\Gamma_2$, $\Omega_R=\Omega/\Gamma_2$, $\omega_{k0}=\omega_k/\Gamma_2$, $\Delta_{pu0}=\Delta_{pu}/\Gamma_2$, $\gamma_{k0}=\gamma_k/\Gamma_2$, $\Gamma_2=\Gamma_1/2$ and $\eta_0=2\sigma_z^0$ is the population inversion of the pseudo-two-level system, which is determined by
\begin{equation}
    \begin{aligned}
        (\eta_0+1)[(\Delta_{pu0}-g^2\omega_{k0}\eta_0)^2+1]+2\Omega_R^2\eta_0=0.
    \end{aligned}
\end{equation}
The linear absorption of the system is proportional to the imaginary part of the first-order optical susceptibility ${\rm Im}(\chi(\omega_{pr}))$. Therefore, it is possible to design an experiment to observe the coupling of the neutron to the GNR through the absorption spectrum of the proposed system. Specifically, one can first drive the entire system with a pump beam, and then we can simply extract the coupling strength between neutrons and GNR from the absorption spectrum of the probe light.

In order to illustrate the feasibility of this experiment, we chose appropriate parameters to demonstrate our proposed scheme. At present, there is no experimental value of $\Gamma_2$, we estimate $\Gamma_2\approx 1.13\times 10^{-12}$~GHz based on the known neutron lifetime $\tau_n^{beam}\approx 888.0$~s. The absorption spectrum of this scheme is shown in Fig.~\ref{fig:wide}. We find that if there is a coupling between neutrons and GNR, a peak will appear at $\delta=\pm \omega_k$ in the absorption spectrum, as shown in Figures Fig.~\ref{fig:wide}(a) ($\delta=-\omega_k$) and \ref{fig:wide}(b) ($\delta=+\omega_k$). Physically, the core reason for this phenomenon is that the coupling of neutrons and GNR induces the splitting of the original energy level of the system. We suggest that the process shown in Fig.~\ref{fig:wide}(a) is due to a three-photon resonance, i.e., the system absorbs two photons of the pump light and then emits one photon, whereas the latter is the ac-Stark shift caused by GNR. We also show the energy level diagrams of corresponding transitions in the inset of Fig.~\ref{fig:wide}(a) and Fig.~\ref{fig:wide}(b).

On the other hand, the height of the both peaks at $\delta=\pm \omega_k$ increases with the enhancement of the neutron-GNR coupling, which is shown in Fig.~\ref{fig:wide}(c). Therefore, the experimental scheme we propose opens up a way to detect whether neutrons couple to GNR (namely trap bottle) and the intensity of the coupling. If the peaks can be observed experimentally, then one can claim that the acceleration of neutron decay induced by the Purcell effect has been achieved in the ongoing neutron decay experiment. In this case, the neutron decay anomaly would be the result of an experimental setup like the bottle-type experiment.

\section{Conclusions} 

In this paper, we have discussed that the Purcell effect can in principle be observed in bottle-type neutron lifetime measurement experiments. In this type of experiment, where ultra-cold neutrons are trapped in a barrier, the quantum effect, in this case the Purcell effect, may become important. Based on the above analysis, we construct a cavity quantum electrodynamics scheme for the neutron-proton state transition, which indicates that the bottle-type experimental setup may have an impact on the neutron lifetime measurement experiments. In addition, using the parameters that can be achieved experimentally, we designed and proposed an experiment that can test whether Purcell effect occurs in future lifetime measurement experiments. Our results show that the current experimental level can detect the Purcell effect that occurs during the neutron lifetime measurement process.

Besides, we have discussed that Purcell effect may have occurred in bottle-type neutron lifetime measurement experiments, and that the so-called "neutron decay anomaly" may be the ultimate result of Purcell effect. If our analysis of the bottle-type neutron lifetime experiment under CQED formalism is correct, then the beam-type experiment measures the true neutron lifetime.

Finally, from a particle physics perspective, as long as the lifetime of neutrons can be further determined, one can explore new physics, because determining the neutron lifetime helps to constrain the measurement of other particles. On the other hand, even though the Purcell effect in neutron decay has not yet been observed, future trap experiments on ultra-cold neutrons could verify this very interesting quantum effect, so it is possible to modify the nuclear decay rate through external experimental environments. This has huge implications for nuclear physics and nuclear engineering.

\begin{acknowledgments}
Fei He would like to thank Jie-Ping Zheng and Zhi-Peng Chen at Shanghai Jiao Tong University for helpful discussion. This work was supported by the National Natural Science Foundation of China (Grants No.11274230 and No.11574206) and Natural Science Foundation of Shanghai (No.20ZR1429900).
\end{acknowledgments}

% The \nocite command causes all entries in a bibliography to be printed out
% whether or not they are actually referenced in the text. This is appropriate
% for the sample file to show the different styles of references, but authors
% most likely will not want to use it.
\nocite{*}

\bibliography{apssamp}% Produces the bibliography via BibTeX.

\end{document}